\documentclass[3p,times,procedia]{elsarticle}
\usepackage{nupha_ecrc}

\volume{00}

\firstpage{1}

\journalname{Nuclear Physics A}

\runauth{}

\jid{nupha}

\jnltitlelogo{Nuclear Physics A}

\usepackage{amssymb}

\usepackage[figuresright]{rotating}

\begin{document}

\begin{frontmatter}



\dochead{XXVIIth International Conference on Ultrarelativistic Nucleus-Nucleus Collisions\\ (Quark Matter 2018)}

\title{Open heavy-flavour production and elliptic flow in p-Pb collisions at the LHC with ALICE}


\author{Henrique J. C. Zanoli (on behalf of the ALICE Collaboration)}

\address{Universidade de S\~ao Paulo, S\~ao Paulo - Brazil \\ Utrecht Universiteit, Utrecht - The Netherlands}

\begin{abstract}
Measurements of open heavy flavour production in p--A collisions allow the investigation of Cold Nuclear Matter effects. In addition, they are an important tool for a complementary investigation of the long-range correlations found in small systems in the light flavour sector. In this work, production measurements of D mesons at mid-rapidity in p--Pb collisions at $\sqrt{s_{\rm NN}}$ = 5.02 TeV are reported. Production yields are also reported for the heavy-flavour hadron decay electrons at central rapidity at $\sqrt{s_{\rm NN}}$ = 5.02 and 8.16 TeV. The elliptic flow ($v_{\rm 2}$) of heavy-flavour hadron decay electrons in high-multiplicity p-Pb collisions at $\sqrt{s_{\rm NN}}$ = 5.02 TeV is found to be positive with a significance larger than $5 \sigma$.

\end{abstract}

\begin{keyword}
heavy-ion collisions \sep heavy-flavour \sep p-Pb collisions \sep ALICE \sep LHC 


\end{keyword}

\end{frontmatter}


\section{Introduction}

Open heavy-flavour measurements in p--A collisions allow the investigation of effects related to the presence of nuclei in the colliding system. 
These include cold-nuclear matter (CNM) effects~\cite{Andronic:2015wma}, the most relevant of which for heavy-flavour particles at LHC energies is shadowing, a reduction of the gluon parton-distribution functions at low Bjorken-\textit{x}. 
Additional possible effects in p--A collisions are the energy loss in CNM, $k_{\rm T}$ broadening, and other initial- or final-state effects. 
They are studied using the nuclear modification factor ($R_{\rm pPb}$), defined as the ratio of the differential cross sections in p--Pb collisions ($\sigma_{\rm pPb}$) divided by the one in pp collisions ($\sigma_{\rm pp}$), normalised by the atomic mass number (X):

\begin{equation}
R_{\rm pPb} = \frac{1}{X} \frac{\rm{d}^2 \sigma_{\rm pPb}/\rm{d} \textit{p}_{\rm T}\rm{d}\textit{y}}{\rm{d}^2 \sigma_{\rm pp}/\rm{d}\textit{p}_{\rm T}\rm{d}\textit{y}}.
\end{equation}
In recent years, long-range $v_{\rm 2}$-like angular correlations, typically observed in nucleus-nucleus collisions and attributed to the collective expansion of the Quark-Gluon Plasma (QGP), have been found in high-multiplicity p--A events. This, along with the observed $v_{\rm 2}$ dependence on particle mass, has motivated to model p--Pb collisions using a medium following a hydrodynamical evolution ~\cite{Werner:2010ss, Deng:2011at}.
The presence of such structure for heavy-flavour particles is still under study. Hence, it is relevant to study p--A collisions to understand possible connections between the mechanisms in A--A and p--A. In this work, the production of D mesons and heavy-flavour hadron decay electrons are reported. More details on the analysis procedures and additional results can be found in Refs.~\cite{ALICE-PUBLIC-2017-008,Acharya:2018dxy}.

\label{sec:intro}

\section{Analysis procedure and data sample}
\label{sec:analysis}
The data sample used for the D meson production studies and the elliptic flow of heavy-flavour decay electrons was collected in 2016 during the LHC p--Pb run at $\sqrt{s_{\rm{NN}}} = 5.02$ TeV with the ALICE detector~\cite{Aamodt:2008zz}. ALICE is located in the Large Hadron Collider in the French-Swiss boarder. The detector has good tracking and PID thanks to the Time Projection Chamber (TPC) and other systems, as described in Ref.~\cite{Abelev:2014ffa}. The events were recorded using a minimum-bias trigger, which required coincident signals in the backward/forward scintillators (V0-A and V0-C arrays). About $6 \times 10^{8}$ events were analysed, corresponding to an integrated luminosity of $L_{\rm{int}} = 295~\pm~11~\mu\rm{b}^{-1}$. The results of the production of heavy-flavour decay electrons are obtained from the analysis of the 2013 and 2016 data samples of p-Pb collisions at $\sqrt{s_{\rm{NN}}} = 5.02$ TeV and $\sqrt{s_{\rm{NN}}}=8.16$ TeV, respectively. 
In this case, an additional trigger for high energy deposition in the electromagnetic calorimeter is employed to extend the measurements towards higher $p_{\rm T}$. 
Events were divided in multiplicity classes using the V0-A detector, in the backward (Pb-going) direction, covering $2.8 < \eta < 5.1$, and in different centrality classes using the ZNA detector (Zero-Degree Neutron Calorimeter), positioned along the beam line in the backward direction. The centrality estimators and its connection with multiplicity classes obtained with various estimator in p-Pb collision are described in Ref.~\cite{Adam:2014qja}. 

D mesons and their charge conjugates were reconstructed in the following hadronic decay channels: $\rm{D^{0}~\rightarrow K^+~\pi^-}$, $\rm{D^+~\rightarrow~K^-~\pi^+ ~\pi^+}$, $\rm{D^{*+} \rightarrow D^{0} \pi^+}$ and $\rm{D^+_s\rightarrow \phi \pi^+ \rightarrow K^- K^+ \pi^+}$. The analysis is based on the identification of secondary vertices with topologies typical of decays displaced from the interaction vertex. The D-meson candidates were reconstructed using pairs or triplets of tracks with the proper charge sign combination. 
Geometrical selections on the D-meson decay topology were applied to reduce the random combinatorial background. Particle identification was also used to further suppress the background. 
Acceptance times efficiency was calculated using Monte Carlo simulations and range from the order of $\approx10^{-1}$\% to $\approx40$\% depending on the transverse momentum and D-meson species. 
For more details about the procedure check Ref.~\cite{ALICE-PUBLIC-2017-008}.

The production of heavy-flavour hadron decay electrons is measured by subtracting to the inclusive electron production the contribution of background electrons. The main background sources are photon conversions ($\gamma \rightarrow \rm{e}^+ \rm{e}^-$) in the beam pipe and in the material of the innermost layers of the ITS and Dalitz decays of neutral mesons ($\rm{\pi}^{0} \rightarrow {\rm \gamma} ~\rm{e}^+ \rm{e}^-$ and ${\rm \eta} \rightarrow {\rm \gamma} ~\rm{e}^+ \rm{e}^-$). The efficiency was also calculated using Monte Carlo simulations and it is approximately 30\%. 
More details can be found in Refs.~\cite{Acharya:2018dxy,Adam:2015qda}.

\section{Results}
\label{sec:results}

\begin{figure}[!t]
\centering
\includegraphics[width= 6.5cm]{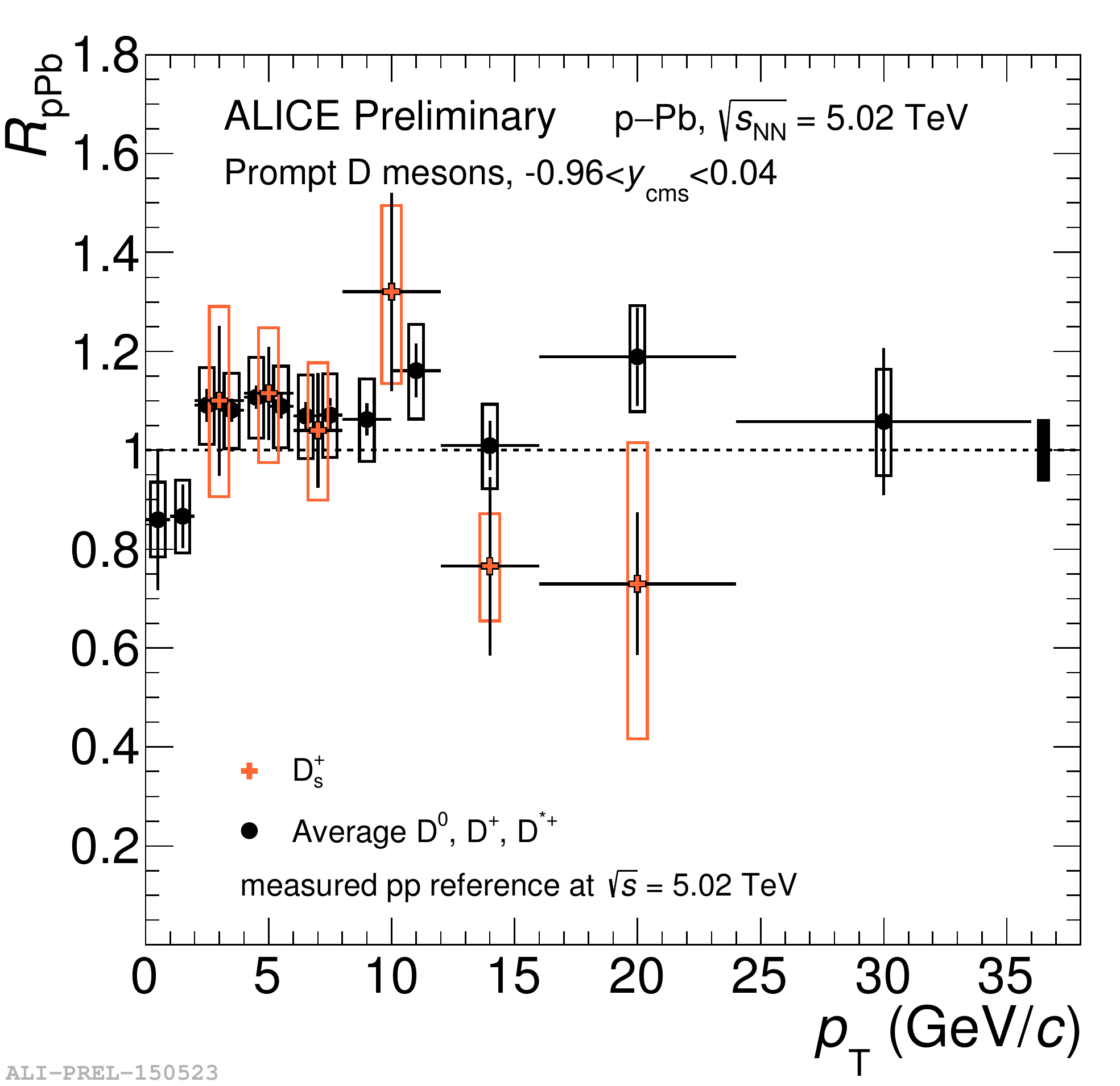}
\includegraphics[width= 6.9 cm]{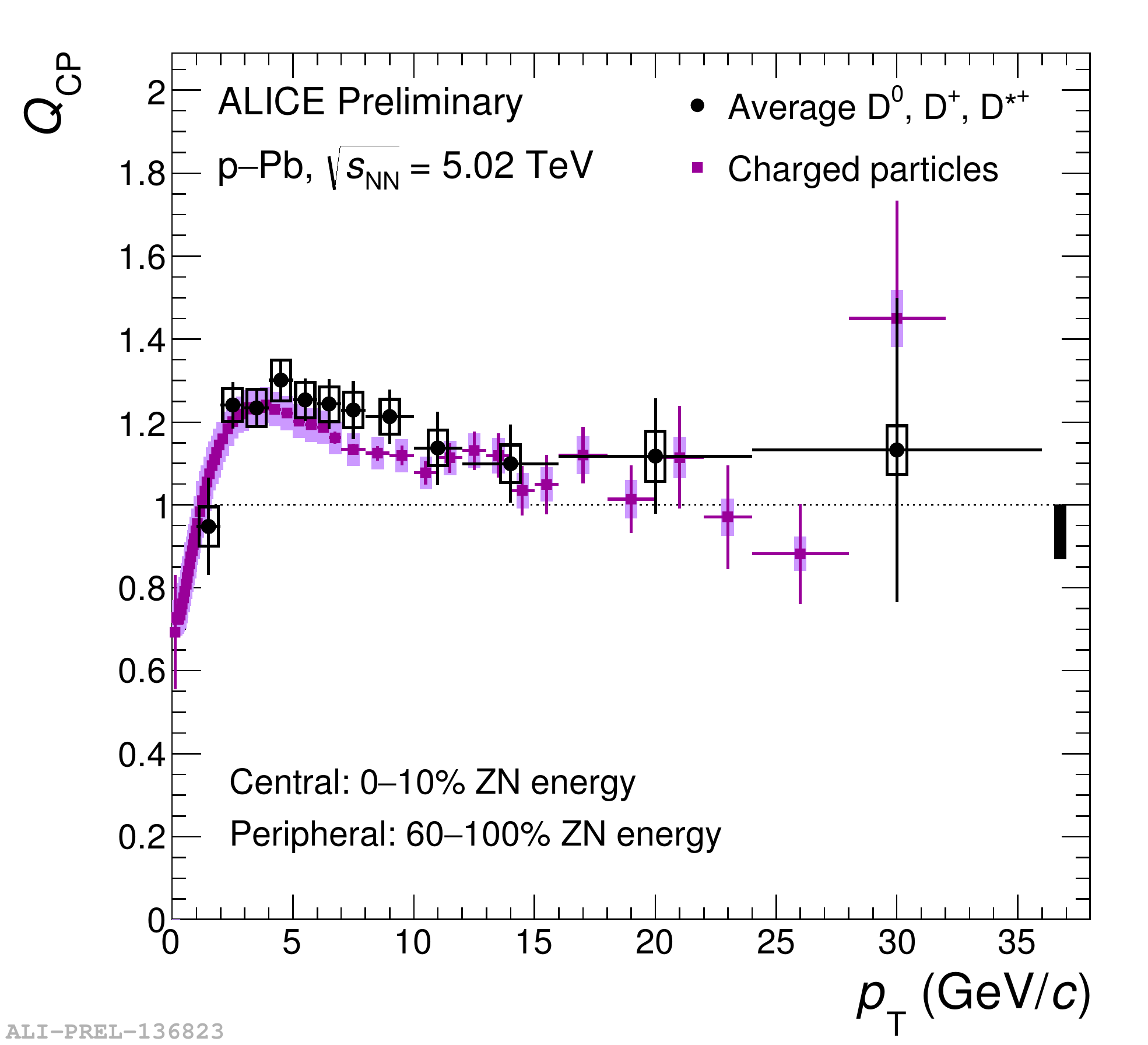}
\caption{Left: nuclear modification factor 
$R_{\rm pPb}$ of prompt D mesons 
in p--Pb collisions at $\sqrt{s_{\rm NN}} =$ 5.02 TeV. Right: average D-meson $Q_{\rm CP}$ for 0-10\%/60-100\% in $1 < p_{\rm T} < 36$ GeV/$c$ compared to charged hadrons $Q_{\rm CP}$ in the same centrality classes. }\label{fig:DRppB}
\end{figure}

The D-meson $p_{\rm T}$ differential $R_{\rm pPb}$ is reported on the left side of Fig. \ref{fig:DRppB}. All the values are compatible with unity within uncertainties. The $R_{\rm pPb}$ was updated using the preliminary measurement of the pp cross section at $\sqrt{s} = 5.02$ TeV~\cite{ALICE-PUBLIC-2018-006}. The new reference reduces the uncertainty and allows to set stringent constraints to models at low transverse momentum. The ${\rm D_s^+}$-meson $R_{\rm pPb}$ is compatible with that of non-strange D mesons. To further study the centrality dependence, the ratio of the yields in central and peripheral collisions was measured, defined as the $Q_{\rm CP}$:

\begin{equation}
Q_{\rm CP} = \frac{ (\rm{d}^2\textit{N} / \rm{d}\textit{p}_{\rm T} \rm{d} \textit{y})^{\rm{central}}_{pPb}  /\langle \textit{T}_{\rm pPb}^{\rm central} \rangle }{(\rm{d}^2\textit{N} / \rm{d}\textit{p}_{\rm T} \rm{d} \textit{y})^{\rm{60-100\%}}_{pPb}  /\langle \textit{T}_{\rm pPb}^{\rm 60-100\%} \rangle },
\end{equation}
where the $(\rm{d}^2\textit{N} / \rm{d}\textit{p}_{\rm T} \rm{d} \textit{y})^{\rm centrality}_{\rm pPb}$ is the yield of the particle of interest in a given centrality and $\langle T_{\rm pPb}^{\rm centrality} \rangle $ is the average nuclear overlapping function for that centrality~\cite{Adam:2014qja}. The results of the $Q_{\rm CP}$ for non-strange D-mesons is shown on the right side of Fig. \ref{fig:DRppB}, compared to the charged particles $Q_{\rm CP}$. They are compatible within uncertainties. The D-meson $Q_{\rm CP}$ for the $p_{\rm T}$ interval $3 < p_{\rm T} < 8 $ GeV/$c$ is larger than unity by 1.5 standard deviations (combining statistical and systematic uncertainties).

\begin{figure}[!b]
\centering
\includegraphics[width= 6.5cm]{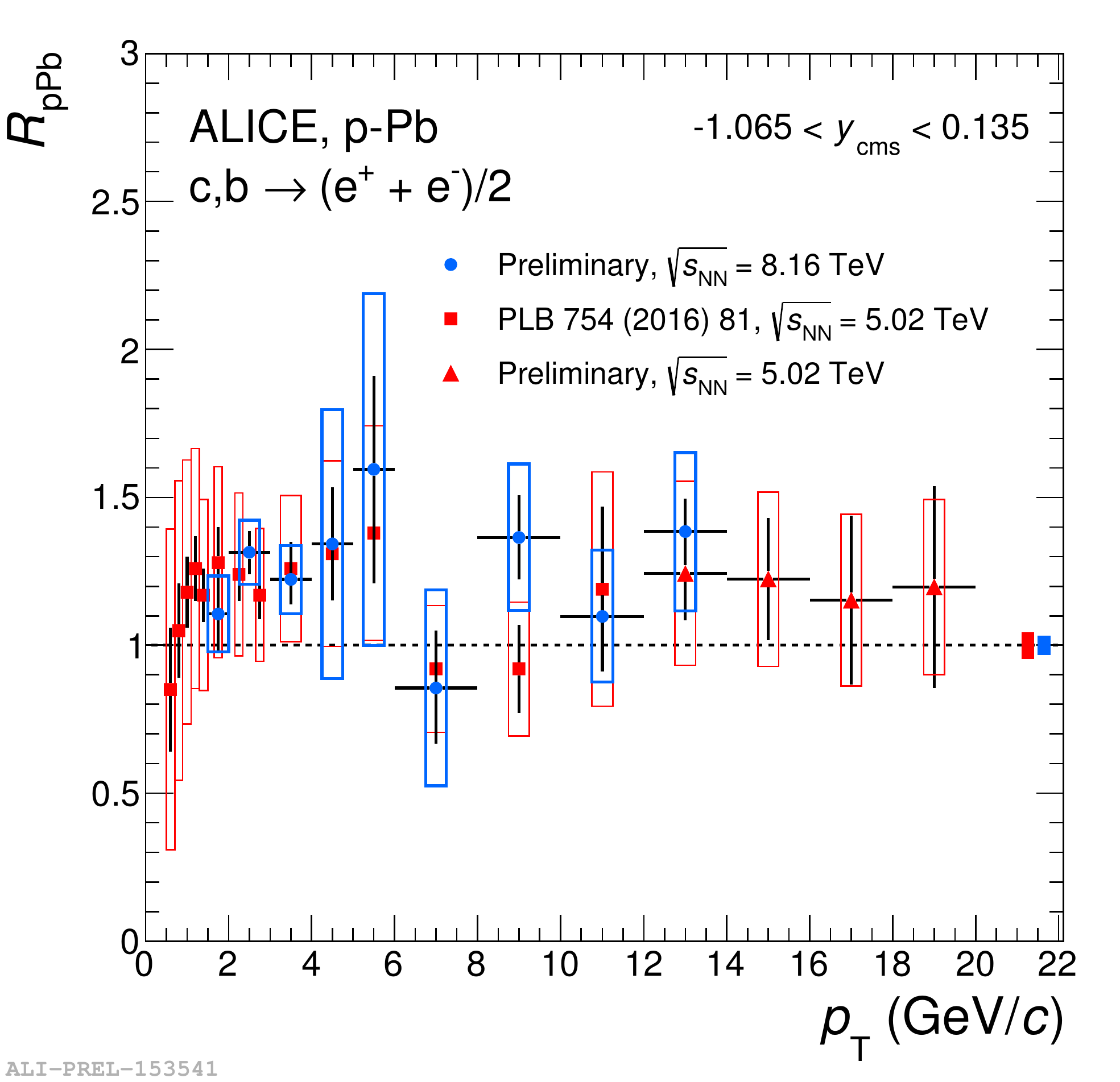}
\caption{Nuclear modification factor 
$R_{\rm pPb}$ of heavy-flavour decay electrons 
in p--Pb collisions at $\sqrt{s_{\rm NN}} =$ 5.02~TeV and 8.16~TeV.}\label{fig:HFeRppB}
\end{figure}

Similar studies are performed for electrons originated from decays of  hadrons that contain heavy quarks \cite{Adam:2015qda} and are presented in Fig. \ref{fig:HFeRppB}. The $p_{\rm T}$ differential heavy-flavour decay electron $R_{\rm pPb}$ is presented for different energies. The uncertainties do not allow to draw conclusions about the energy dependence. The measurements are compatible with unity in all the $p_{\rm T}$ interval studied.

\begin{figure}[!t]
\centering
\includegraphics[width= 7.25cm]{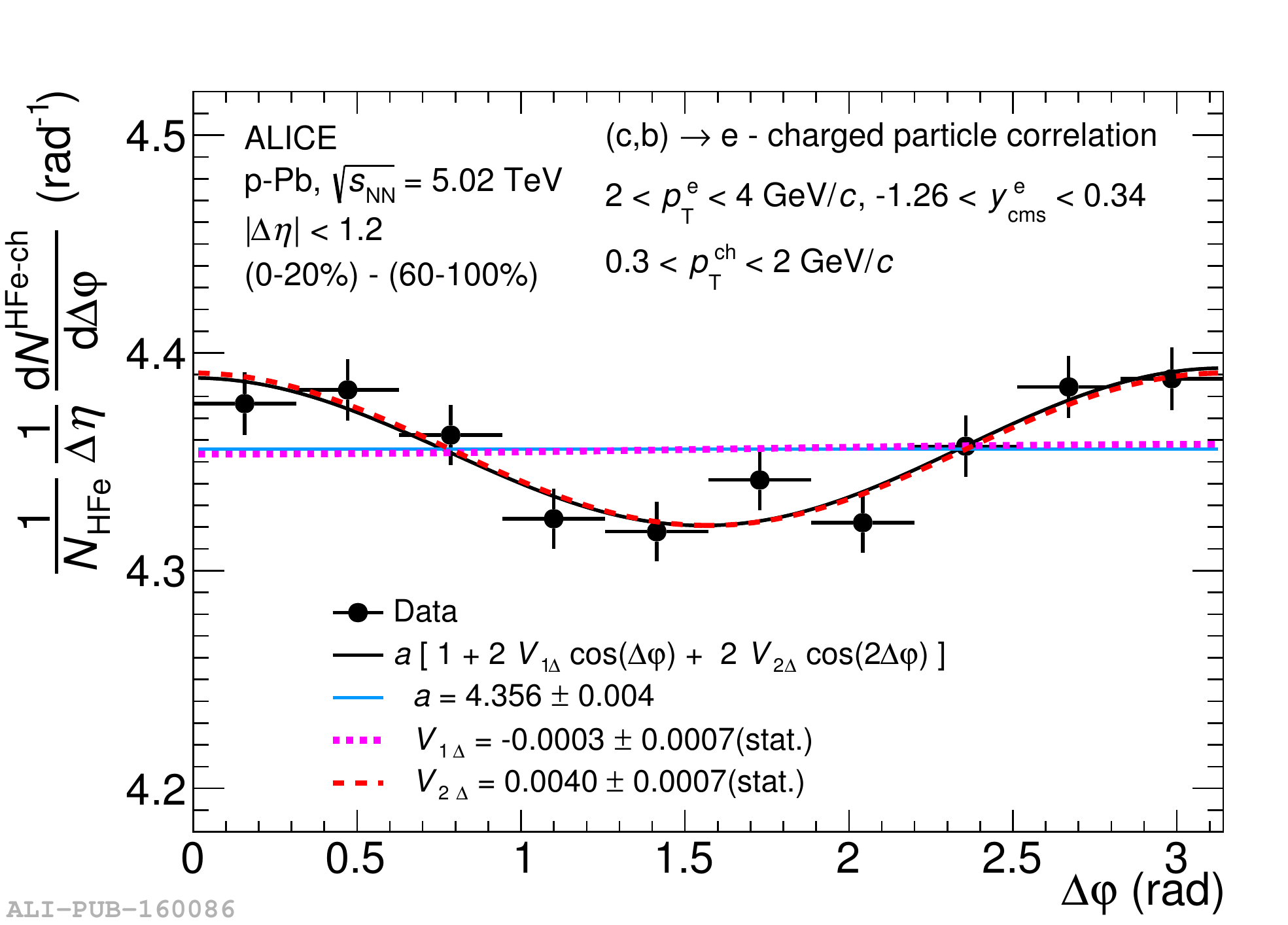}
\includegraphics[width= 7.25cm]{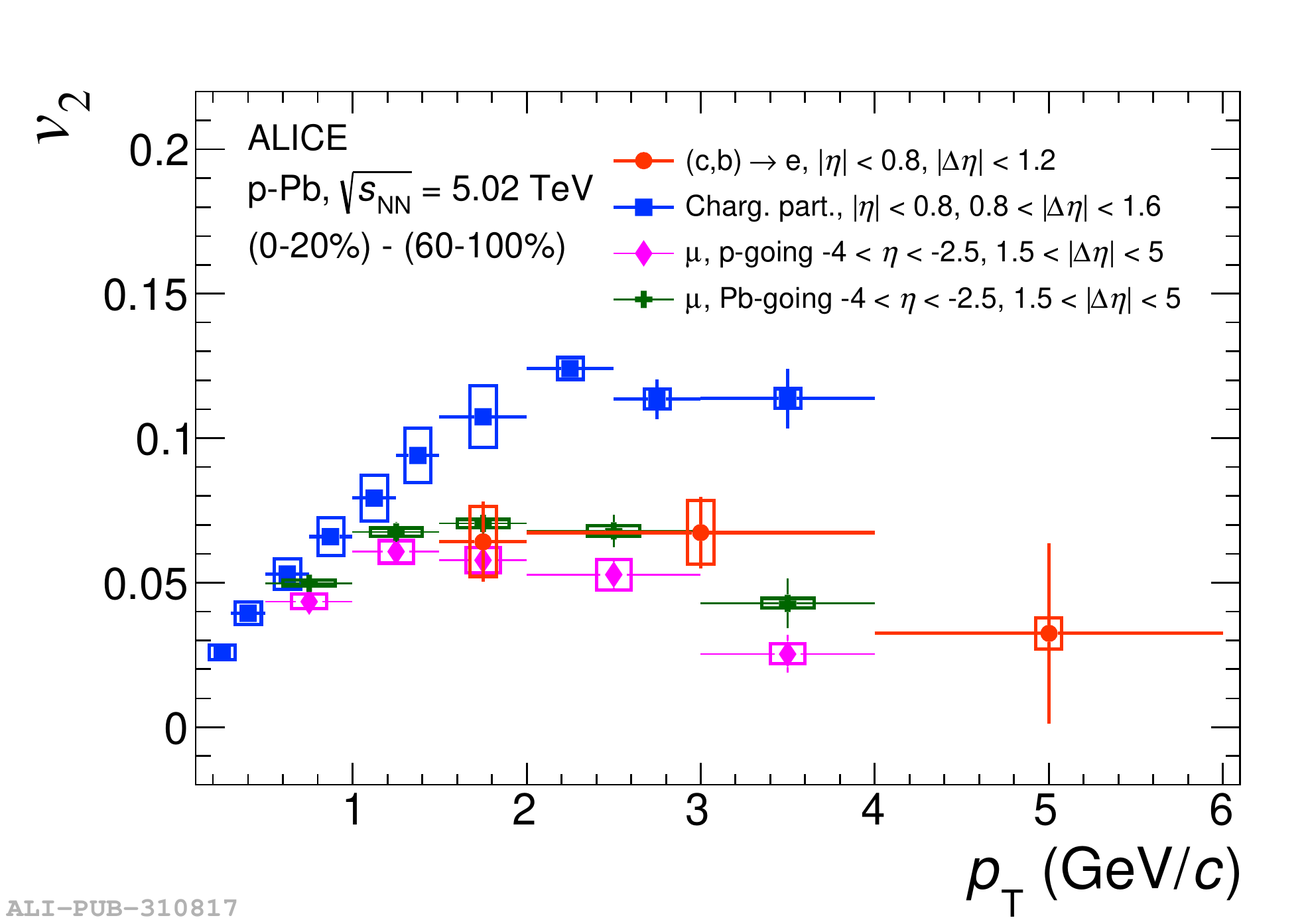}
\caption{Left: azimuthal-correlation distribution between heavy-flavour decay electrons and charged particles, for high-multiplicity p--Pb collisions after subtracting the jet contribution from low-multiplicity collisions. Right panel: heavy-flavour decay electron $v_2$ as a function of $p_{\rm T}$ compared to the $v_2$ of unidentified charged particles~\cite{ABELEV:2013wsa} and inclusive muons \cite{Adam:2015bka}.}\label{fig:HFev2}
\end{figure}

The results of the study of the angular correlations between heavy-flavour decay electrons and charged particles in high multiplicity collisions are reported on Fig. \ref{fig:HFev2}. The analysis is performed in two different multiplicity classes: 0-20\% (high multiplicity) and 60-100\% (low multiplicity), obtained using the V0A detector. After the removal of the jet contribution to the correlation function (left side of Fig. \ref{fig:HFev2}), using the correlation distribution  obtained in the low multiplicity class,
a $v_{2}$-like modulation was observed in the high-multiplicity correlation distributions, similarly to what was previously observed for light-flavour di-hadron distributions~\cite{ABELEV:2013wsa}. 
A Fourier fit to the correlation distributions was used to characterise the modulation and extract the heavy-flavour electron $v_{2}$. The heavy-flavour decay electron $v_{2}$ (right side of  Fig. \ref{fig:HFev2}) is positive with a significance of 5.1$\sigma$ in the $1.5 < p_{\rm T}^{\rm e} < 4$ GeV/$c$ interval.

\section{Summary}
\label{sec:summary}
We have presented the $R_{\rm pPb}$ of D mesons in p--Pb collisions at $\sqrt{s_{\rm NN}}$ = 5.02 TeV and of heavy-flavour hadron decay electrons in p--Pb collisions at $\sqrt{s_{\rm NN}}$ = 5.02 TeV and 8.16 TeV. They are compatible with unity, indicating that the suppression observed in Pb-Pb collisions is probably due to final-state effects. The new preliminary measurement of D-meson $R_{\rm pPb}$ has smaller uncertainties than previous measurements thanks to the new pp reference at 5.02 TeV~\citep{ALICE-PUBLIC-2018-006}. The ratio of the production of D-mesons in central and peripheral collisions ($Q_{\rm CP}$) for $3 < p_{\rm{T}} < 8 $ GeV/$c$ is larger than unity by 1.5 standard deviations. The study of the two-particle correlation between heavy-flavour decay electrons and charged particles shows evidence of a positive $v_2$ coefficient with a significance of more than 5$\sigma$ for $1.5 < p_{\rm T} < 4$ GeV/c. 

\section*{Acknowledgements}
H. Zanoli thanks the financial support of FAPESP (grants 2014/11042-9 and 2016/05723-9) .




\bibliographystyle{elsarticle-num}
\bibliography{bib.bib}







\end{document}